\documentclass[english]{article}
\usepackage[T1]{fontenc}
\usepackage{textcomp}
\usepackage[latin9]{inputenc}
\usepackage{mathrsfs}
\usepackage{amsmath}
\usepackage{amssymb}

\makeatletter
\newcommand{\lyxaddress}[1]{
	\par {\raggedright #1
	\vspace{1.4em}
	\noindent\par}
}

\usepackage{fullpage}

\makeatother

\usepackage{babel}
\begin{document}
\title{Poincare gauge gravity from nonmetric gravity}
\author{James T. Wheeler$^{\dagger}$}
\maketitle
\begin{abstract}
We consider general linear gauge theory, with independent solder form
and connection. These spaces have both torsion and nonmetricity. We
show that the Cartan structure equations together with the defining
equation for nonmetricity allow the mixed symmetry components of nonmetricity
to be absorbed into an altered torsion tensor. Field redefinitions
reduce the structure equations to those of Poincare gauge theory,
with local Lorentz symmetry and metric compatibility.

In order to allow recovery the original torsion and nonmetric fields,
we replace the definition of nonmetricity by an additional structure
equation and demand integrability of the extended system. We show
that the maximal Lie algebra compatible with the enlarged set is isomorphic
to the conformal Lie algebra. From this Lorentzian conformal geometry,
we establish that the difference between the field strength of special
conformal transformations and the torsion and is given by the mixed
symmetry nonmetricity of an equivalent asymmetric system.
\end{abstract}

\lyxaddress{Keywords: Poincare gauge theory, Netric-affine gravity, General linear
gauge theory, Torsion, Nonmetricity, Nonmetric gravity, General relativity,
Conformal, Biconformal, Scale invariant general relativity}

$^{\dagger}$James T Wheeler, Utah State University Department of
Physics, 4415 Old Main Hill, Logan, UT 84322-4415, jim.wheeler@usu.edu

\newpage{}

\section{Introduction}

Poincaré gauge theory places general relativity in a gauge theory
context similar to other fundamental interactions. There are important
differences from other gauge theories, however, since the gravity
connection arises from the metric. This additional field leads to
the natural inclusion of torsion in addition to the curvature as descriptive
of the geometry. Assuming vanishing torsion and the Einstein-Hilbert
action we recover general relativity, but there are reasons to prefer
a more general context.

The Poincaré symmetry has two Casimir operators, that is, combinations
of the group generators that are invariant under the full group action.
These Casimir operators correspond to the mass and spin of particle
states. When we include matter sources and vary the solder form and
spin connection independently, these properties (in the form of the
energy tensor and spin density) give sources for the curvature and
torsion respectively. Despite the absence of any direct experimental
evidence for torsion, this shows a tight connection between the elements
of Poincaré gauge theory and properties of physical fields. This motivates
exploration of Poincaré gauge theory in greater detail.

Here we approach Poincaré gauge theory from the other direction, studying
a broader class of gravity theories and showing that many reduce to
Poincaré gauge theory or its conformal extension. By dropping any
a priori restriction on the connection, we leave open the question
of whether the metric and connection are compatible. If compatibility
is not assumed, the gauge theory will differ from canonical general
relativity by the presence of both torsion and nonmetricity. Nonmetricity--i.e.,
the covariant derivative of the metric--depends on a symmetric contribution
to the spin connection, and this is of particular concern because
any symmetric part of the spin connection breaks Lorentz invariance.

To explore these issues, our current study begins with a general linear
gauge theory. The resulting general form of the connection means that
nearly any choice of metric will produce nonmetricity, with no particular
metric preferred. We choose a Lorentzian metric since we want to see
how Poincaré symmetry lies within this broader context. We find that
although the general linear theory develops nonmetricity, the mixed-symmetry
part of the nonmetricity may always be eliminated by field redefinitions.
Using the nonmetricity to antisymmetrize the spin connection restores
a Lorentz connection. This gives an explicit realization of a Lorentzian
signature extension of Theorem 5.8 of \cite{Isham}. Within the recast
theory, field redefinitions merge the mixed-symmetry nonmetricity
and torsion into a single modified torsion field, eliminating their
independence and restoring the appearance of Poincaré gauge theory.

The field redefinitions mask the original form of the connection,
leaving torsion and part of the nonmetricity inextricably merged.
We develop sufficient additional structure to recover the original
nonmetricity and torsion separately. Introducing an additional structure
equation to characterize the nonmetricity, we use integrability to
fix the maximal extension of Poincaré symmetry that includes the new
field. We find this maximal extension of the nonmetric structure is
equivalent to a gauge theory of the conformal group, and shows a surprising
equivalence between nonmetricity, torsion, and the special conformal
curvature.

Throughout, we work in arbitrary dimension $d$ using Cartan methods
\cite{Collected Cartan,Kobayashi and Nomizu,Neeman and Regge} to
build fiber bundles of specific symmetry. This gives the arena for
all gauged theories of those symmetries.

The progression of our investigation is as follows. In the next Section,
we develop the Cartan equations for general linear symmetry, and show
that field redefinitions restore the Poincaré form of the structure
equations. Redefinition of the curvature eliminates any remaining
evidence of nonmetricity. Lorentz covariance is restored while independence
of the original fields is lost.

In Section (\ref{sec:Restoring-independence-of}) we add an additional
structure equation to the modified Lorentzian system in such a way
that the original fields may be recovered. This involves the introduction
of a potential for the mixed-symmetry part of the nonmetricity. We
check integrability of the extended system, and backtrack to find
an equivalent extended nonmetric system.

The addition of an additional variable in Section (\ref{sec:Restoring-independence-of})
ignores the fact that the new potential might also contribute to the
other structure equations. In Section (\ref{sec:The-maximal-Lie})
we establish a maximal Cartan system by adding a linear combination
of all allowed terms that are at least linear in the new potential
to the nonmetric structure equations of Section (\ref{sec:Restoring-independence-of}).
We determine the coefficients and ensure consistency by demanding
that the enlarged system arises from the Maurer-Cartan equations of
some Lie algebra. This is accomplished by setting curvature, torsion,
and nonmetricity to zero, then imposing integrability of the full
set to fix the coefficients. The remaining equations are then necessarily
the Maurer-Cartan equations of a Lie algebra. Redefining fields to
produce a Lorentzian system, we show that the maximal system is equivalent
to the conformal Maurer-Cartan equations. 

The extended Lie algebra may then be used to define a gravity theory.
This extends the original asymmetric system to a maximal set that,
when gauged in Section (\ref{sec:Developing-the-curvatures}) to include
the field strengths, allows us to determine both the original torsion
and nonmetricity. Using basis changes we relate the nonmetric curvature,
the torsion and the nonmetricity to the curvature, torsion, and special
conformal curvature of the conformal theory.

All calculations to this point are based on the structure equations
and not on any particular physical theory, and apply in any dimension
$d>3$. In the final Section, summarize our findings and briefly comment
on two conformal gravity theories. We show that in the auxiliary conformal
gauge gravity there is a class of solutions in which the field strength
of the special conformal gauge field explicitly becomes the nonmetricity.
In the biconformal, or Kähler, gauging the interchangeability of the
torsion and the mixed symmetry nonmetricity is manifest.

The remaining totally symmetric part of the nonmetricity is likely
to vanish without the--possibly unlikely \cite{de Wit Freedman,Stackexchange}--inclusion
of spin-3 sources. In the absence of spin-3 sources we may conclude
that the study of gauge theories of gravity with general connection
may be recast as the study of conformal gravity theories.

\section{Poincaré gauge theory from general linear gauge theory}

We wish to form a principal fiber bundle with $GL\left(d,\mathbb{R}\right)$
symmetry group and base manifold $\mathcal{M}^{\left(d\right)}$.
To achieve this we start with the inhomogeneous group $IGL\left(d,\mathbb{R}\right)$.
With restriction to $d=4$, the basic structures agree with those
found in studies \cite{HehlKerlickVDHeyde,IvanenkoSardanashvily,HehlMetricAffine}
of the metric-affine theory of gauged $GL\left(4,\mathbb{R}\right)$.

\subsection{Manifold with local $GL\left(d,\mathbb{R}\right)$ fiber bundle}

We write a $\left(d+1\right)$-dimensional representation to find
the Lie algebra. Expanding near the identity, the extra column and
row permit infinitesimal translations,
\[
\left(\begin{array}{cc}
1+\left[\varepsilon_{ij}\right]_{\;\;\;n}^{m} & a^{m}\\
0 & 1
\end{array}\right)\left(\begin{array}{c}
x^{n}\\
1
\end{array}\right)=\left(\begin{array}{c}
x^{m}+\left[\varepsilon_{ij}\right]_{\;\;\;n}^{m}x^{n}+a^{m}\\
1
\end{array}\right)
\]
We identify generators $E_{ij\;\;\;\;N}^{\;\;M}=\left(\begin{array}{cc}
\left[\varepsilon_{ij}\right]_{\;\;\;n}^{m} & a^{m}\delta_{n}^{5}\\
0 & 0
\end{array}\right)$, where $\left[\varepsilon_{ij}\right]_{\;\;\;n}^{m}$ generate $GL\left(d,\mathbb{R}\right)$.
For a basis of $GL\left(d,\mathbb{R}\right)$ generators let $\left[\varepsilon_{ij}\right]_{\;\;\;m}^{k}=\delta_{i}^{k}\eta_{jm}$
where $\eta_{jm}$ is chosen to be the Lorentzian $SO\left(d-1,1\right)$
metric. This allows us to relate the geometry to a Lorentzian spacetime,
even though there is no a priori metric for the general linear space.
The translation generators $\lambda_{k\;\;\;\;N}^{\;\;M}=\left(\begin{array}{cc}
0 & \delta_{k}^{m}\delta_{n}^{5}\\
0 & 0
\end{array}\right)$ lead to additive group transformations $T_{\;\;\;K}^{M}\left(a^{k}\right)T_{\;\;\;N}^{K}\left(b^{k}\right)=T_{\;\;\;N}^{M}\left(a^{k}+b^{k}\right)$
where $T_{\;\;\;N}^{M}\left(a^{m}\right)=\left(\begin{array}{cc}
\delta_{n}^{m} & a^{m}\delta_{n}^{5}\\
0 & 1
\end{array}\right)$.

The $\mathfrak{igl}\left(d\right)$ Lie algebra follows immediately
as
\begin{eqnarray*}
\left[\varepsilon_{ij},\varepsilon_{mn}\right] & = & c_{\quad ij,mn}^{kl}\varepsilon_{kl}\\
\left[\varepsilon_{ij},\lambda_{k}\right] & = & c_{\;\;\;ij,k}^{m}\lambda_{m}\\
\left[\lambda_{i},\lambda_{j}\right] & = & 0
\end{eqnarray*}
where the structure constants are $c_{\quad ij,mn}^{kl}=\eta_{in}\delta_{m}^{k}\delta_{j}^{l}-\eta_{mj}\delta_{i}^{k}\delta_{n}^{l}$
and $c_{\;\;\;ij,k}^{m}=\eta_{jk}\delta_{i}^{m}$.

Let $\left\langle \varepsilon_{ij},\boldsymbol{\alpha}^{mn}\right\rangle =\delta_{i}^{m}\delta_{j}^{n}$
and $\left\langle \lambda_{i},\mathbf{e}^{j}\right\rangle =\delta_{i}^{j}$
define the dual 1-forms. Then the Maurer-Cartan structure equations
for $\mathfrak{igl}\left(n\right)$ are
\begin{eqnarray*}
\mathbf{d}\boldsymbol{\alpha}_{\;\;\;j}^{i} & = & \boldsymbol{\alpha}_{\;\;\;j}^{k}\wedge\boldsymbol{\alpha}_{\;\;\;k}^{i}\\
\mathbf{d}\mathbf{e}^{i} & = & \mathbf{e}^{k}\wedge\boldsymbol{\alpha}_{\;\;\;k}^{i}
\end{eqnarray*}

The quotient of the inhomogeneous general linear group, $IGL\left(d\right)$,
by its general linear subgroup
\[
\mathcal{M}_{0}^{\left(d\right)}=IGL\left(d\right)/GL\left(d\right)
\]
is a homogeneous $d$-dimensional manifold. Defining a projection
$\pi:H_{g}\rightarrow\mathcal{M}_{0}^{\left(d\right)}$ that maps
the cosets of the quotient to this manifold we produce a principal
fiber bundle, which we generalize by changing the connection, $\boldsymbol{\alpha}_{\;\;\;j}^{i}\rightarrow\boldsymbol{\beta}_{\;\;\;j}^{i}$.
This introduces a curvature 2-form into each equation,
\begin{eqnarray}
\mathbf{d}\boldsymbol{\beta}_{\;\;\;j}^{i} & = & \boldsymbol{\beta}_{\;\;\;j}^{n}\wedge\boldsymbol{\beta}_{\;\;\;n}^{i}+\boldsymbol{\mathcal{R}}_{\;\;\;j}^{i}\label{GL curvature}\\
\mathbf{d}\mathbf{e}^{i} & = & \mathbf{e}^{k}\wedge\boldsymbol{\beta}_{\;\;\;k}^{i}+\boldsymbol{\mathcal{T}}^{i}\label{GL torsion}
\end{eqnarray}
These 2-forms are each tensorial under the fiber symmetry $GL\left(d,\mathbb{R}\right)$.
To maintain the principal fiber bundle we require $\boldsymbol{\mathcal{R}}_{\;\;\;j}^{i}$
and $\boldsymbol{\mathcal{T}}^{i}$ to be horizontal in the bundle,
\begin{eqnarray*}
\boldsymbol{\mathcal{R}}_{\;\;\;j}^{i} & = & \frac{1}{2}\mathcal{R}_{\;\;\;jkl}^{i}\mathbf{e}^{k}\wedge\mathbf{e}^{l}\\
\boldsymbol{\mathcal{T}}^{i} & = & \frac{1}{2}\mathcal{T}_{\;\;\;kl}^{i}\mathbf{e}^{k}\wedge\mathbf{e}^{l}
\end{eqnarray*}
At the same time, if we choose, we may change the base manifold, $\mathcal{M}_{0}^{\left(d\right)}\rightarrow\mathcal{M}^{\left(d\right)}$.

\subsection{Nonmetricity}

We allow the solder forms $\mathbf{e}^{i}$ to describe an inverse
orthonormal metric
\begin{eqnarray*}
\left\langle \mathbf{e}^{i},\mathbf{e}^{j}\right\rangle  & = & \eta^{ij}
\end{eqnarray*}
of Lorentz signature $\left(d-1,1\right)$. In the linear theory this
is an arbitrary choice which is justified because it is not required
to be compatible with the connection. Moreover, it is necessary in
order to separate the symmetric (nonmetric) and antisymmetric (Lorentzian)
parts of the connection. Therefore, we have nonmetricity
\begin{eqnarray*}
\mathbf{Q}_{ij} & \equiv & \mathbf{d}\eta_{ij}-\eta_{kj}\boldsymbol{\beta}_{\;\;\;i}^{k}-\eta_{ik}\boldsymbol{\beta}_{\;\;\;j}^{k}\\
 & = & -\boldsymbol{\beta}_{ji}-\boldsymbol{\beta}_{ij}
\end{eqnarray*}
from which the symmetric part of the connection is
\begin{eqnarray}
\boldsymbol{\beta}_{\left(ij\right)} & = & -\frac{1}{2}\mathbf{Q}_{ij}\label{Symmetric as Q}
\end{eqnarray}
Note that this form is independent of the signature of the orthonormal
metric\footnote{We may write the group generators as $\left[\varepsilon^{ij}\right]_{km}=\delta_{k}^{i}\delta_{m}^{j}$
so the doubly covariant form of the connection $\alpha_{ij\mu}\left[\varepsilon^{ij}\right]_{kl}$
is independent of the orthonormal form of the metric $\eta_{mn}$.
Therefore, the nonmetricity given by the symmetric part $Q_{km\mu}=-2\alpha_{ij\mu}\left[\varepsilon^{ij}\right]_{\left(km\right)}$
is independent of the choice of orthonormal metric. Once we choose
the signature, no alternative choice of metric can give zero nonmetricity,
since nonmetricity is a tensor. However, since $\mathbf{d}g_{ij}=\left(Q_{ij\alpha}+\alpha_{ij\mu}+\alpha_{ji\alpha}\right)u^{\alpha}\mathbf{d}\lambda$
is integrable along any timelike congruence of curves $u^{\alpha}$,
we may find $g_{ij}$ such that $Q_{ij\alpha}u^{\alpha}=0$ along
the congruence.}.

With the symmetric part of the connection given by (\ref{Symmetric as Q}),
we may separate the structure equations into symmetric and antisymmetric
parts. Let
\begin{equation}
\boldsymbol{\beta}_{\;\;\;j}^{i}=\boldsymbol{\omega}_{\;\;\;j}^{i}-\frac{1}{2}\mathbf{Q}_{\;\;\;j}^{i}\label{GL connection in terms of Lorentz}
\end{equation}
where $\boldsymbol{\omega}_{ij}=-\boldsymbol{\omega}_{ji}$ is a Lorentz
connection. We replace the full asymmetric connection $\boldsymbol{\beta}_{\;\;\;j}^{i}$
in the structure equations by substituting (\ref{GL connection in terms of Lorentz})
into Eqs.(\ref{GL curvature}) and (\ref{GL torsion}), This yields
\begin{eqnarray}
\mathbf{d}\boldsymbol{\omega}_{\;\;\;j}^{i} & = & \boldsymbol{\omega}_{\;\;\;j}^{k}\wedge\boldsymbol{\omega}_{\;\;\;k}^{i}+\boldsymbol{\mathcal{R}}_{\;\;\;j}^{i}+\frac{1}{2}\mathbf{D}\mathbf{Q}_{\;\;\;j}^{i}+\frac{1}{4}\mathbf{Q}_{\;\;\;j}^{k}\wedge\mathbf{Q}_{\;\;\;k}^{i}\label{GL curvature with connection separated}\\
\mathbf{d}\mathbf{e}^{i} & = & \mathbf{e}^{k}\wedge\boldsymbol{\omega}_{\;\;\;k}^{i}+\boldsymbol{\mathcal{T}}^{i}-\mathbf{Q}^{i}\label{GL torsion with connection separated}
\end{eqnarray}
where $\mathbf{D}$ is the Lorentz covariant exterior derivative and
we define the mixed-symmetry part of the nonmetricity as a 2-form
\begin{eqnarray*}
\mathbf{Q}^{i} & \equiv & \frac{1}{2}\mathbf{e}^{k}\wedge\mathbf{Q}_{\;\;\;k}^{i}\\
 & = & \frac{1}{2}Q_{\;\;\;\left[km\right]}^{i}\mathbf{e}^{k}\wedge\mathbf{e}^{m}
\end{eqnarray*}
Decomposing the nonmetricity as $Q_{ijk}=Q_{\left(ijk\right)}+\frac{2}{3}\left(Q_{\left[ij\right]k}+Q_{\left[ik\right]j}\right)$
reveals its two irreducible subspaces spanned by $Q_{\left(ijk\right)}$
and $Q_{i\left[jk\right]}$, so that $\mathbf{Q}^{i}$ is the larger
of the irreducible pieces ($\frac{1}{3}d\left(d^{2}-1\right)\geq\frac{1}{6}d\left(d+1\right)\left(d+2\right)$
when $d>3$).

Separating $\boldsymbol{\mathcal{R}}_{\;\;\;j}^{i}$ into symmetric
$\check{\boldsymbol{\mathcal{R}}}_{ij}=\boldsymbol{\mathcal{R}}_{\left(ij\right)}$
and antisymmetric $\hat{\boldsymbol{\mathcal{R}}}_{ij}=\boldsymbol{\mathcal{R}}_{\left[ij\right]}$
parts, we have two independent equations. Identifying the Lorentz
forms of the curvature $\mathbf{R}_{\;\;\;j}^{i}=\mathbf{d}\boldsymbol{\omega}_{\;\;\;j}^{i}-\boldsymbol{\omega}_{\;\;\;j}^{k}\wedge\boldsymbol{\omega}_{\;\;\;k}^{i}$
and torsion $\mathbf{T}^{i}=\mathbf{D}\mathbf{e}^{i}$, we make the
field redefinitions
\begin{eqnarray}
\mathbf{R}_{\;\;\;j}^{i} & \equiv & \hat{\boldsymbol{\mathcal{R}}}_{\;\;\;j}^{i}-\frac{1}{4}\mathbf{Q}_{\;\;\;k}^{i}\wedge\mathbf{Q}_{\;\;\;j}^{k}\label{Redefined curvature}\\
\mathbf{T}^{i} & \equiv & \boldsymbol{\mathcal{T}}^{i}-\mathbf{Q}^{i}\label{Redefined torsion}
\end{eqnarray}
in Eqs.(\ref{GL curvature with connection separated}) and (\ref{GL torsion with connection separated})
to form the Cartan equations of Poincaré gauge theory
\begin{eqnarray}
\mathbf{d}\boldsymbol{\omega}_{\;\;\;j}^{i} & = & \boldsymbol{\omega}_{\;\;\;j}^{k}\wedge\boldsymbol{\omega}_{\;\;\;k}^{j}+\mathbf{R}_{\;\;\;j}^{i}\label{Lorentz curvature}\\
\mathbf{d}\mathbf{e}^{i} & = & \mathbf{e}^{k}\wedge\boldsymbol{\omega}_{\;\;\;k}^{i}+\mathbf{T}^{i}\label{Lorentz torsion}
\end{eqnarray}
with an additional symmetric field $\check{\boldsymbol{\mathcal{R}}}_{ij}$
given by
\begin{eqnarray*}
\check{\boldsymbol{\mathcal{R}}}_{ij} & = & -\frac{1}{2}\mathbf{D}\mathbf{Q}_{ij}
\end{eqnarray*}
The nonmetricity of the Lorentz connection $\boldsymbol{\omega}_{\;\;\;j}^{i}$
is, of course, zero. The symmetric part of the curvature may be rewritten
as two covariant exterior derivatives acting on the metric.
\[
\check{\boldsymbol{\mathcal{R}}}_{ij}=-\frac{1}{2}\mathbf{D}\mathbf{D}\mathbf{\eta}_{ij}=-\frac{1}{4}\mathbf{d}x^{\alpha}\wedge\mathbf{d}x^{\beta}\left[D_{\alpha},D_{\beta}\right]g_{\mu\nu}
\]
Working out the commutator, the symmetric combination of Lorentz curvatures
drops out and we are left with
\begin{eqnarray*}
\check{\boldsymbol{\mathcal{R}}}_{\mu\nu} & = & -\frac{1}{2}\mathbf{T}^{\sigma}Q_{\mu\nu\sigma}
\end{eqnarray*}
so the symmetric part of the curvature can be expressed as a contraction
of torsion with nonmetricity. This last expression has vanishing Ricci
scalar, so it does not contribute to the Einstein-Hilbert action.

The replacement of Eqs.(\ref{GL curvature with connection separated})
and (\ref{GL torsion with connection separated}) by Eqs.(\ref{Lorentz curvature})
and (\ref{Lorentz torsion}) is our first principal result: \emph{Field
redefinitions allow us to write the Cartan equations of general linear
gauge theory as Poincaré gauge theory, with the mixed symmetry part
of nonmetricity combining as an altered torsion.}

\smallskip{}

Identifying $\mathbf{Q}^{i}$ as torsion-like goes far beyond previous
results. A very limited overlap between one contraction of the nonmetricity
and the trace of the torsion was shown by Smalley \cite{Smalley}
in 1986, and noted many times since in the context of various generalized
geometries \cite{Obukhov,KarananasMonin,IosifidisPetkouTsagas,KlemmRavera,SauroMartiniZanusso,Condeescu}.
This relation is easy to see since the solder form structure equation
of a Weyl geometry may be written either by (1) including the Weyl
vector $\boldsymbol{\omega}=W_{i}\mathbf{e}^{i}$ as a nonmetric contribution
\begin{eqnarray*}
\mathbf{d}\mathbf{e}^{i} & = & \mathbf{e}^{k}\wedge\boldsymbol{\omega}_{\;\;\;k}^{i}+\boldsymbol{\omega}\wedge\mathbf{e}^{i}+\mathbf{T}^{i}\\
 & = & \mathbf{e}^{k}\wedge\left(\boldsymbol{\omega}_{\;\;\;k}^{i}-\boldsymbol{\omega}\delta_{\;\;\;k}^{i}\right)+\mathbf{T}^{i}
\end{eqnarray*}
with $\mathbf{Q}_{ij}=\mathbf{D}\eta_{ij}=2\eta_{ij}\boldsymbol{\omega}$,
or (2) as a redefined torsion
\begin{eqnarray*}
\mathbf{d}\mathbf{e}^{i} & = & \mathbf{e}^{k}\wedge\boldsymbol{\omega}_{\;\;\;k}^{i}+\boldsymbol{\omega}\wedge\mathbf{e}^{i}+\mathbf{T}^{i}\\
 & = & \mathbf{e}^{k}\wedge\boldsymbol{\omega}_{\;\;\;k}^{i}+\frac{1}{2}\left(T_{\;\;\;jk}^{i}+\delta_{\;\;\;j}^{i}W_{k}-\delta_{\;\;\;k}^{i}W_{j}\right)\mathbf{e}^{j}\wedge\mathbf{e}^{i}
\end{eqnarray*}
This relation only applies to a single vector. By contrast, the mixing
of half or more ($d>3$) of the degrees of freedom of nonmetricity
into the modified torsion seen in Eq.(\ref{Redefined torsion}) goes
far beyond the Smalley ambiguity in the interpretation of the Weyl
vector, $W_{i}$.

\subsection{Sources with the linear connection}

Next, consider the gravitational field equations. Properties of nonmetricity
have received considerable attention starting from early work \cite{HehlKerlickVDHeyde,IvanenkoSardanashvily,HehlMetricAffine}
and progressing to greatly expanded interest in recent years (e.g.,
\cite{SauroMartiniZanusso}, \cite{Recent 01}-\cite{Recent 15}).
Here we briefly summarize some of the known results, and comment on
the vacuum case.

We work with Eqs. (\ref{GL curvature with connection separated})
and (\ref{GL torsion with connection separated}) with the original
fields, varying the metric, torsion, and nonmetricity independently.
Writing the action in the Einstein-Hilbert form $\int\boldsymbol{\mathcal{R}}^{ij}\wedge\mathbf{e}^{k}\wedge\mathbf{e}^{l}e_{ijkl}$
plus a source term $S_{source}=\int\boldsymbol{\mathcal{L}}_{source}$,
the $\mathbf{D}\mathbf{Q}_{\;\;\;j}^{i}$ term in Eq.(\ref{GL curvature with connection separated})
drops out by symmetry. The remaining scalar curvature becomes $R=\hat{\mathcal{R}}-\frac{1}{8}\left(Q_{\;\;\;mk}^{k}Q_{\;\;\;\;\;i}^{mi}-Q^{kmi}Q_{mik}\right)$,
where the scalar $\hat{\mathcal{R}}$ still depends on the modified
torsion $\mathbf{T}^{i}\equiv\boldsymbol{\mathcal{T}}^{i}-\mathbf{Q}^{i}$.
The metric and torsion variations follow the usual patterns from general
relativity and ECSK theory. Here we limit our attention to sources
provided by the Standard Model, and to the vacuum case.

Varying nonmetricity, the general form of the field equation is
\begin{eqnarray}
\frac{1}{8}\left(Q_{cba}+Q_{cab}-\eta_{ac}Q_{\;\;\;be}^{e}-\eta_{bc}Q_{\;\;\;ae}^{e}\right) & = & -\frac{\delta\mathcal{L}_{Source}}{\delta Q^{abc}}\label{Field equation for nonmetricity}
\end{eqnarray}
where the source on the right has been called the hypermomentum \cite{IosifidisHehl,ObukhovHehl2024b,HehlEtAlHypermomentum2,HehlEtAlHypermomentum3,ObukhovHehl2024a}.

There are three types of matter fields within the Standard Model:
the Higgs scalar doublet, $U\left(1\right)$ and Yang-Mills fields,
and Dirac spinors. The covariant derivative of scalar fields in curved
spacetime do not require the spacetime connection (although the Higgs
invariance requires the $SU\left(2\right)$ connection). The same
is true of Yang-Mills fields, for which the derivatives are spacetime
curls, again requiring only the internal symmetry connection but not
the spacetime connection. Therefore, varying the part of the general
linear connection dependent on the nonmetricity will not produce any
scalar or Yang-Mills couplings.

Coupling to Dirac spinors does involve the spacetime spin connection.
But as observed in \cite{HehlEtAlHypermomentum2,HehlEtAlHypermomentum3},
the real-valued group $GL\left(4,\mathbb{R}\right)$ cannot contain
the $Spin\left(3,1\right)$ representation given by Dirac spinors\footnote{$GL\left(d,\mathbb{R}\right)$ does contain Spin$\left(p,q\right)$
in certain other dimensions and signatures}. Therefore, before we can understand Dirac coupling to nonmetricity,
we must generalize the Lorentzian spin connection. We have carried
this out with the result that Dirac fields \emph{do} produce nonmetricity.
However discussing this here would take us too far afield. For this
reason, we postpone a detailed treatment of Dirac couplings to a separate
study \cite{Wheeler 2025b}.

\subsubsection{Vacuum nonmetricity from Einstein-Hilbert}

In vacuum the source on the right side of Eq.(\ref{Field equation for nonmetricity})
vanishes and we have
\begin{eqnarray*}
Q_{c\left(ab\right)}-\eta_{(a\left|c\right|}Q_{\;\;\;b)e}^{e} & = & 0
\end{eqnarray*}
Contraction shows that $Q_{\;\;\;ae}^{e}=\frac{1}{d}Q_{\;\;\;ea}^{e}$.
The latter contraction is twice the Weyl vector, $W_{a}=\frac{1}{2d}Q_{\;\;\;ea}^{e}$
so the symmetric part of the nonmetricity is $Q_{c\left(ab\right)}=\eta_{ac}W_{b}+\eta_{ab}W_{c}$
and the full nonmetricity can be expressed in terms of $\boldsymbol{\omega}$
and $\mathbf{Q}^{a}$ only.

\section{Restoring independence of the nonmetricity \label{sec:Restoring-independence-of}}

We now turn to our second principal result. The form of Eqs.(\ref{Lorentz curvature})
and (\ref{Lorentz torsion}) confounds any independent prediction
of nonmetricity and torsion, at least in the vacuum theory. Only the
emergent final forms $\mathbf{R}_{\;\;\;b}^{a},\mathbf{T}^{a}$ enter
the structure equations, and if the action is built from these, the
field equations determine only the modified curvature and torsion.
Now we develop an additional equation to describe nonmetricity. This
extends the independent variables from $\left(\boldsymbol{\omega}_{\;\;\;b}^{a},\mathbf{e}^{a},\boldsymbol{\omega}\right)$
to $\left(\boldsymbol{\omega}_{\;\;\;b}^{a},\mathbf{e}^{a},\mathbf{f}_{a},\boldsymbol{\omega}\right)$
so that solving for all three connection forms determines the curvature,
torsion \emph{and} the nonmetricity. With the addition of this single
new structure equation, which we call the minimal extension, the full
set remains integrable. However introducing a new field might also
modify the original structure equations as well. In Section (\ref{sec:The-maximal-Lie})
we form the maximal extension of the $\left(\boldsymbol{\omega}_{\;\;\;b}^{a},\mathbf{e}^{a},\mathbf{f}_{a},\boldsymbol{\omega}\right)$
system by including all such modifications. Integrability of the resulting
maximal system results in the Maurer-Cartan equations of the conformal
group.

\subsection{Evaporating nonmetricity with separated Weyl vector}

We begin again with Eqs.(\ref{GL curvature with connection separated})-(\ref{GL torsion with connection separated}),
this time separating the Weyl vector, which we include in the antisymmetric
part of the connection. To check this, solve for the connection in
the usual way. Separate the nonmetricity into trace and traceless
parts,
\begin{eqnarray*}
Q_{abc} & = & 2\eta_{ab}W_{c}+\tilde{Q}_{abc}
\end{eqnarray*}
and define the altered torsion using the original \emph{traceless}
nonmetricity $\tilde{\mathbf{Q}}^{a}$ only, $\mathbf{T}^{a}\equiv\boldsymbol{\mathcal{T}}^{a}-\tilde{\mathbf{Q}}^{a}$.
Substituting $\hat{\boldsymbol{\omega}}_{\;\;\;b}^{a}=\boldsymbol{\omega}_{\;\;\;b}^{a}-\frac{1}{2}\left(\tilde{\mathbf{Q}}_{\;\;\;b}^{a}+2\delta_{b}^{a}W_{c}\mathbf{e}^{c}\right)$
into the solder form equation (\ref{GL torsion}) results in a Weyl
connection, which then enters the curvature (\ref{GL curvature})
in the usual way. The modified form of the structure equations is
now,

\begin{eqnarray}
\mathbf{d}\boldsymbol{\omega}_{\;\;\;b}^{a} & = & \boldsymbol{\omega}_{\;\;\;b}^{c}\wedge\boldsymbol{\omega}_{\;\;\;c}^{a}+\mathbf{R}_{\;\;\;b}^{a}\label{Weyl geometry curvature}\\
\mathbf{d}\mathbf{e}^{a} & = & \mathbf{e}^{b}\wedge\boldsymbol{\omega}_{\;\;\;b}^{a}+\boldsymbol{\omega}\wedge\mathbf{e}^{a}+\mathbf{T}^{a}\label{Weyl geometry torsion}\\
\mathbf{d}\boldsymbol{\omega} & = & \boldsymbol{\Omega}\label{Weyl vector}
\end{eqnarray}
with $\boldsymbol{\omega}=W_{c}\mathbf{e}^{c}$. Here the traceless
nonmetricity vanishes from the formulation. Equations (\ref{Weyl geometry curvature})-(\ref{Weyl vector})
describe a Weyl geometry with Weyl vector $\boldsymbol{\omega}$ and
torsion $\mathbf{T}^{a}$. Including the Weyl vector in the structure
equation for the solder form does not change its antisymmetry, $\boldsymbol{\omega}_{ab}=-\boldsymbol{\omega}_{ba}$,
so that we still have a Lorentzian spacetime. Agreement with local
scale invariance (i.e., changes of units) requires the Weyl vector
to be integrable so that $\boldsymbol{\Omega}=0$. This condition
frequently follows from the field equations. 

As in the previous Section, all reference to the original nonmetricity
has vanished and the remaining spin connection is Lorentzian.

\subsection{Independent nonmetricity and torsion}

The torsion equation, Eq.(\ref{Weyl geometry torsion}) now determines
only one combination of the original two fields $\boldsymbol{\mathcal{T}}^{a},\tilde{\mathbf{Q}}^{a}$.
If we wish to recover both original fields we need an independent
equation. The equation for $\mathbf{h}_{a}=g_{ab}\mathbf{e}^{b}$
explored in \cite{Wheeler 2023b} is suggestive, but results in $\mathbf{d}\mathbf{h}_{a}=\boldsymbol{\omega}_{\;\;\;a}^{d}\wedge\mathbf{h}_{d}+\mathbf{h}_{a}\wedge\boldsymbol{\omega}+\boldsymbol{\mathcal{T}}^{a}-\tilde{\mathbf{Q}}^{a}$,
which is just Eq.(\ref{Weyl geometry torsion}) written in covariant
form.

To form an independent equation, we write a similar equation,
\begin{eqnarray}
\mathbf{d}\mathbf{f}_{a} & = & \boldsymbol{\omega}_{\;\;\;a}^{b}\wedge\mathbf{f}_{b}+\mathbf{f}_{a}\wedge\boldsymbol{\omega}+\bar{\mathbf{T}}_{a}\label{Nonmetricity structure eq}
\end{eqnarray}
where we change the field strength to $\bar{\mathbf{T}}_{a}\equiv\boldsymbol{\mathcal{T}}_{a}+\tilde{\mathbf{Q}}_{a}$.
The 1-form $\mathbf{f}_{a}$ is now independent of the solder form
so that from an appropriate action the combined system Eqs.(\ref{Weyl geometry curvature})-(\ref{Nonmetricity structure eq})
can determine both $\boldsymbol{\mathcal{T}}_{a}$ and $\tilde{\mathbf{Q}}_{a}$.

The nonmetricity vanishes if $\mathbf{f}_{a}=\mathbf{h}_{a}$, so
the sum and difference 1-forms
\begin{eqnarray}
\mathbf{u}^{a} & \equiv & \frac{1}{2}\left(\mathbf{e}^{a}+g^{ab}\mathbf{f}_{b}\right)\nonumber \\
\mathbf{v}_{a} & \equiv & \frac{1}{2}\left(\mathbf{f}_{a}-g_{ab}\mathbf{e}^{b}\right)\label{(u,v) in terms of (e,f) basis}
\end{eqnarray}
give separate equations for $\boldsymbol{\mathcal{T}}^{a}$ and $\tilde{\mathbf{Q}}_{a}$
respectively. Simplifying the expressions for $\mathbf{d}\mathbf{u}^{a}$
and $\mathbf{d}\mathbf{v}_{a}$ and noting that $\mathbf{d}g_{ab}=\boldsymbol{\omega}_{ab}+\boldsymbol{\omega}_{ba}-2g_{ab}\boldsymbol{\omega}=-2g_{ab}\boldsymbol{\omega}$
we now have the extended collection:
\begin{eqnarray}
\mathbf{d}\boldsymbol{\omega}_{\;\;\;b}^{a} & = & \boldsymbol{\omega}_{\;\;\;b}^{c}\wedge\boldsymbol{\omega}_{\;\;\;c}^{a}+\mathbf{R}_{\;\;\;b}^{a}\nonumber \\
\mathbf{d}\mathbf{u}^{a} & = & \mathbf{u}^{b}\wedge\boldsymbol{\omega}_{\;\;\;b}^{a}+\boldsymbol{\omega}\wedge\mathbf{u}^{a}+\boldsymbol{\mathcal{T}}^{a}\nonumber \\
\mathbf{d}\mathbf{v}_{a} & = & \boldsymbol{\omega}_{\;\;\;a}^{b}\wedge\mathbf{v}_{b}+\mathbf{v}_{a}\wedge\boldsymbol{\omega}+\tilde{\mathbf{Q}}_{a}\nonumber \\
\mathbf{d}\boldsymbol{\omega} & = & \boldsymbol{\Omega}\label{Cartan Eqs so far}
\end{eqnarray}
where
\begin{eqnarray*}
\mathbf{R}_{\;\;\;b}^{a} & \equiv & \boldsymbol{\mathcal{R}}_{\;\;\;b}^{a}-\frac{1}{2}\mathbf{D}\tilde{\mathbf{Q}}_{\;\;\;b}^{a}+\frac{1}{4}\tilde{\mathbf{Q}}_{\;\;\;b}^{c}\wedge\tilde{\mathbf{Q}}_{\;\;\;c}^{a}
\end{eqnarray*}
As a check, when we set $\mathbf{v}_{a}=0$ and gauge the Weyl vector
to zero we return to the original recovered Poincaré theory, Eqs.(\ref{Lorentz curvature})
and (\ref{Lorentz torsion}).

The full system now has field strengths equal to the original torsion
and traceless nonmetricity of the asymmetric connection. 

Retracing the calculation backwards from Eqs.(\ref{Cartan Eqs so far}),
we find that the equations in the $\left(\mathbf{e}^{a},\mathbf{f}_{a}\right)$
basis written with the original asymmetric (i.e., general linear)
connection take the form
\begin{eqnarray}
\mathbf{d}\hat{\boldsymbol{\omega}}_{\;\;\;b}^{a} & = & \hat{\boldsymbol{\omega}}_{\;\;\;b}^{c}\wedge\hat{\boldsymbol{\omega}}_{\;\;\;c}^{a}+\boldsymbol{\mathcal{R}}_{\;\;\;b}^{a}\nonumber \\
\mathbf{d}\mathbf{e}^{a} & = & \mathbf{e}^{b}\wedge\hat{\boldsymbol{\omega}}_{\;\;\;b}^{a}+\boldsymbol{\mathcal{T}}^{a}\nonumber \\
\mathbf{d}\mathbf{f}_{a} & = & \hat{\boldsymbol{\omega}}_{\;\;\;a}^{b}\wedge\mathbf{f}_{b}+\boldsymbol{\mathcal{K}}_{a}\nonumber \\
\mathbf{d}\boldsymbol{\omega} & = & \boldsymbol{\Omega}\label{Asymmetric Cartan equations}
\end{eqnarray}
where
\begin{eqnarray*}
\boldsymbol{\mathcal{K}}_{a} & \equiv & \boldsymbol{\mathcal{T}}_{a}+\frac{1}{2}\left(\mathbf{e}^{b}-g^{bc}\mathbf{f}_{c}\right)\wedge\tilde{\mathbf{Q}}_{ab}
\end{eqnarray*}
As expected, the dependence on nonmetricity is proportional to the
difference $\mathbf{e}^{b}-g^{bc}\mathbf{f}_{c}$, so that when $\mathbf{f}_{a}=g_{ab}\mathbf{e}^{b}$
the nonmetricity equation simply replicates the torsion equation.
We have then recovered the original nonmetric system.

Next, we examine consistency of the new set of equations.

\subsection{Integrability of nonmetric system \label{subsec:Integrability-of-nonmetric}}

Other than the parallel structure, we have no reason to expect that
including Eq.(\ref{Nonmetricity structure eq}) with the original
Cartan equations (\ref{Weyl geometry curvature})-(\ref{Weyl vector})
gives a consistent set of equations. When the curvature, torsion,
and nonmetricity all vanish the resulting vacuum equations must be
integrable. Generalized Bianchi identities then extend integrability
to the full curved geometry.

Combined integrability of the four equations (\ref{Weyl geometry curvature})-(\ref{Nonmetricity structure eq})
guarantees that the corresponding dual vectors will satisfy closed
commutation relations and the Jacobi identity. Then the extended equations
must be the Maurer-Cartan equations of some Lie algebra.

Integrability is immediate, since we know that with vanishing curvature
and torsion Eqs.(\ref{MC Lorentz})-(\ref{Maurer Cartan dilatation})
comprise the Maurer-Cartan equation of the Weyl group. For the $\mathbf{f}_{a}$
equation we easily check that $\mathbf{d}^{2}\mathbf{f}_{a}\equiv0$
and integrability is established. We note that the proofs do not depend
on the antisymmetry of the spin connection.

This shows that Eqs.(\ref{Weyl geometry curvature})-(\ref{Nonmetricity structure eq})
are consistent, but they do not include possible couplings between
the new field and the original equations. If we regard the vacuum
form of the original equations (\ref{Weyl geometry curvature})-(\ref{Weyl vector})
as those describing some larger Lie algebra with $\mathbf{f}_{a}$
set to zero then we must consider possible $\mathbf{f}_{a}$-dependent
terms in the original three equations as well. In the next Section
we establish the existence of that larger algebra by (1) adding general
$\mathbf{f}_{a}$-dependent terms to each of the original three equations,
(2) adjoining the fourth equation for $\mathbf{f}_{a}$, and finally
(3) demanding integrability to fix the remaining constants. This completes
the extension by giving the Maurer-Cartan equations for some Lie algebra.

Before finding this maximal extension, we solve for the homogeneous
manifold of the minimal set, described by the Maurer-Cartan equations
\begin{eqnarray}
\mathbf{d}\boldsymbol{\omega}_{\;\;\;b}^{a} & = & \boldsymbol{\omega}_{\;\;\;b}^{c}\wedge\boldsymbol{\omega}_{\;\;\;c}^{a}\label{MC Lorentz}\\
\mathbf{d}\mathbf{e}^{a} & = & \mathbf{e}^{b}\wedge\boldsymbol{\omega}_{\;\;\;b}^{a}+\boldsymbol{\omega}\wedge\mathbf{e}^{a}\label{MC translations}\\
\mathbf{d}\mathbf{f}_{a} & = & \boldsymbol{\omega}_{\;\;\;a}^{b}\wedge\mathbf{f}_{b}+\mathbf{f}_{a}\wedge\boldsymbol{\omega}\label{MC New translation}\\
\mathbf{d}\boldsymbol{\omega} & = & 0\label{Maurer Cartan dilatation}
\end{eqnarray}
Even though this does not yet give the maximal form of the Lie algebra,
we gain some insight. 

The Lie group described by Eqs.(\ref{MC Lorentz})-(\ref{Maurer Cartan dilatation})
is not hard to identify: in addition to Lorentz transformations and
dilatations, we now have two sets of translational gauge fields. As
proof, note that Eq.(\ref{MC Lorentz}) shows the spin connection
to be pure (local Lorentz) gauge, $\boldsymbol{\omega}_{\;\;\;b}^{a}=-\mathbf{d}\Lambda_{\;\;\;c}^{a}\bar{\Lambda}_{\;\;\;b}^{c}$.
Choosing a cross section with constant $\Lambda_{\;\;\;b}^{a}$ we
have vanishing spin connection, $\boldsymbol{\omega}_{\;\;\;b}^{a}=0$.
A parallel argument holds for the Weyl vector, so we gauge to $\boldsymbol{\omega}=0$.
The remaining equations are simply
\begin{eqnarray*}
\mathbf{d}\mathbf{e}^{a} & = & 0\\
\mathbf{d}\mathbf{f}_{a} & = & 0
\end{eqnarray*}
and we conclude the existence of functions $x^{a},y_{a}$ such that
\begin{eqnarray*}
\mathbf{e}^{a} & = & \mathbf{d}x^{a}\\
\mathbf{f}_{a} & = & \mathbf{d}y_{a}
\end{eqnarray*}
We assume that $\mathbf{e}^{a}$ is the usual solder form, and exact
orthonormal frames describe flat spaces. Therefore, with the maximal
extension of the coordinates $x^{a}$, the $\mathbf{d}x^{a}$ span
$d$-dimensional Minkowski space with $\mathbf{e}^{a}$ dual to the
generator of translations. While $\mathbf{f}_{a}=\mathbf{d}y_{a}$
may be degenerate or not, the maximal extension of $y_{a}$ in $\mathbb{R}^{d}$
gives two Lorentz covariant possibilities, depending on whether the
$y_{a}$ coordinates are independent of the $x^{a}$:
\begin{enumerate}
\item The potential $\mathbf{f}_{a}$ is linearly dependent on the solder
form, $\mathbf{f}_{a}=b_{ab}\left(x^{c}\right)\mathbf{e}^{b}$.
\item The $\mathbf{f}_{a}$ are independent of the $\mathbf{e}^{a}$, so
that $\left(\mathbf{e}^{a},\mathbf{f}_{b}\right)$ span a flat, $2d$-dimensional
space comprised of two copies of Minkowski space. 
\end{enumerate}
We describe these two cases in more detail in our discussion Section. 

\section{The maximal Lie algebra \label{sec:The-maximal-Lie}}

We found equations (\ref{MC Lorentz})-(\ref{Maurer Cartan dilatation})
by introducing an equation for a new gauge potential $\mathbf{f}_{a}$.
Now we allow the new potential to enter the original equations. Keeping
Eq.(\ref{MC New translation}) for $\mathbf{d}\mathbf{f}_{a}$, but
allowing an asymmetric connection, we seek the maximal form of the
Lie algebra by including $\mathbf{f}_{a}$-dependent terms in the
remaining structure equations. We then limit the coefficients by enforcing
integrability.

We begin with the original general linear geometry, Eqs.(\ref{Asymmetric Cartan equations}),
separate the trace of the nonmetricity to include the Weyl vector
explicitly, then set the curvatures to zero to study the original
vacuum algebra. Appending all possible additional terms to Eqs.(\ref{MC Lorentz})-(\ref{Maurer Cartan dilatation})
we write

\begin{eqnarray}
\mathbf{d}\hat{\boldsymbol{\omega}}_{\;\;\;b}^{a} & = & \hat{\boldsymbol{\omega}}_{\;\;\;b}^{c}\wedge\hat{\boldsymbol{\omega}}_{\;\;\;c}^{a}+\boldsymbol{\Lambda}_{\;\;\;b}^{a\;\;\quad c}\land\mathbf{f}_{c}\nonumber \\
\mathbf{d}\mathbf{e}^{a} & = & \mathbf{e}^{b}\wedge\hat{\boldsymbol{\omega}}_{\;\;\;b}^{a}+\boldsymbol{\omega}\wedge\mathbf{e}^{a}+\boldsymbol{\Lambda}^{ac}\land\mathbf{f}_{c}\nonumber \\
\mathbf{d}\mathbf{f}_{a} & = & \hat{\boldsymbol{\omega}}_{\;\;\;a}^{b}\wedge\mathbf{f}_{b}+\mathbf{f}_{a}\wedge\boldsymbol{\omega}\nonumber \\
\mathbf{d}\boldsymbol{\omega} & = & \boldsymbol{\Lambda}^{c}\land\mathbf{f}_{c}\label{Maximal extensionj}
\end{eqnarray}
where the 1-forms $\boldsymbol{\Lambda}_{\;\;\;b}^{a\quad c},\boldsymbol{\Lambda}^{ac},\boldsymbol{\Lambda}^{c}$
must be built from the available tensors, $g_{ab},\delta_{b}^{a},g^{ab},e_{abcd},e^{abcd}$
and gauge fields $\left(\boldsymbol{\omega}_{\;\;\;b}^{a},\mathbf{e}^{a},\mathbf{f}_{a},\boldsymbol{\omega}\right)$. 

With the curvatures vanishing we have $\mathbf{Q}_{ab}=0$ and therefore
the general linear connection, $\hat{\boldsymbol{\omega}}_{\;\;\;b}^{a}=\boldsymbol{\omega}_{\;\;\;b}^{a}-\frac{1}{2}\mathbf{Q}_{\;\;\;b}^{a}$
reduces to its Lorentzian part, $\boldsymbol{\omega}_{\;\;\;b}^{a}$.
Then, with $\boldsymbol{\Lambda}_{\;\;\;b}^{a\quad c}$ antisymmetric
on $ab$, the possible expressions for the 1-forms of $\boldsymbol{\Lambda}_{\;\;\;b}^{a\quad c},\boldsymbol{\Lambda}^{ac},\boldsymbol{\Lambda}^{c}$
are
\begin{eqnarray*}
\boldsymbol{\Lambda}_{\;\;\;b}^{a\quad c} & = & \alpha e_{\;\;\;b\;\;\quad d}^{a\;\;\quad c}\mathbf{e}^{d}+\beta\left(\delta_{b}^{c}\delta_{d}^{a}-g^{ac}g_{bd}\right)\mathbf{e}^{d}+\mu\left(g^{ad}\delta_{b}^{c}-g^{ac}\delta_{b}^{d}\right)\mathbf{f}_{d}\\
\boldsymbol{\Lambda}^{ac} & = & \rho\boldsymbol{\omega}^{ac}+\lambda e_{\;\quad bd}^{ac}\boldsymbol{\omega}^{bd}+\sigma g^{ac}\boldsymbol{\omega}\\
\boldsymbol{\Lambda}^{c} & = & \gamma\mathbf{e}^{c}+\nu g^{cd}\mathbf{f}_{d}
\end{eqnarray*}
with arbitrary constants $\alpha,\beta,\mu,\rho,\lambda,\sigma,\gamma,\nu$.
Substituting, we demand integrability of
\begin{eqnarray}
\mathbf{d}\boldsymbol{\omega}_{\;\;\;b}^{a} & = & \boldsymbol{\omega}_{\;\;\;b}^{c}\wedge\boldsymbol{\omega}_{\;\;\;c}^{a}+\alpha e_{\;\;\;b\;\;\quad d}^{a\;\;\quad c}\mathbf{e}^{d}\land\mathbf{f}_{c}+\beta\left(\delta_{b}^{c}\delta_{d}^{a}-g^{ac}g_{bd}\right)\mathbf{e}^{d}\land\mathbf{f}_{c}+\mu\left(g^{ad}\delta_{b}^{c}-g^{ac}\delta_{b}^{d}\right)\mathbf{f}_{d}\land\mathbf{f}_{c}\label{Extended connection}\\
\mathbf{d}\mathbf{e}^{a} & = & \mathbf{e}^{b}\wedge\boldsymbol{\omega}_{\;\;\;b}^{a}+\boldsymbol{\omega}\wedge\mathbf{e}^{a}+\rho\boldsymbol{\omega}^{ac}\land\mathbf{f}_{c}+\lambda e_{\;\quad bd}^{ac}\boldsymbol{\omega}^{bd}\land\mathbf{f}_{c}+\sigma g^{ac}\boldsymbol{\omega}\land\mathbf{f}_{c}\label{Extended solder form}\\
\mathbf{d}\mathbf{f}_{a} & = & \boldsymbol{\omega}_{\;\;\;a}^{b}\wedge\mathbf{f}_{b}+\mathbf{f}_{a}\wedge\boldsymbol{\omega}\label{Extended f}\\
\mathbf{d}\boldsymbol{\omega} & = & \left(\gamma\mathbf{e}^{c}+\nu g^{cd}\mathbf{f}_{d}\right)\land\mathbf{f}_{c}=\gamma\mathbf{e}^{c}\land\mathbf{f}_{c}\label{Extended Weyl}
\end{eqnarray}

Starting with the Weyl vector, we see that $\nu$ does not contribute,
so we drop it. Then integrability requires
\begin{eqnarray*}
0 & = & \mathbf{d}^{2}\boldsymbol{\omega}\\
 & = & \gamma\mathbf{d}\mathbf{e}^{a}\land\mathbf{f}_{a}-\gamma\mathbf{e}^{a}\land\mathbf{d}\mathbf{f}_{a}
\end{eqnarray*}
Substituting Eqs.(\ref{Extended solder form}) and (\ref{Extended f})
and collecting terms, we find
\begin{eqnarray}
0 & = & \gamma\left(\rho\hat{\boldsymbol{\omega}}^{ac}+\lambda e_{\;\quad bd}^{ac}\hat{\boldsymbol{\omega}}^{bd}+\sigma g^{ac}\boldsymbol{\omega}\right)\land\mathbf{f}_{c}\land\mathbf{f}_{a}\label{Weyl integrability}
\end{eqnarray}
In general, in order to establish independence of two different terms,
we need only imagine some particular form of the basis forms that
makes them distinct. Also notice that the dual of a 2-form cannot
cancel with the 2-form unless the connection is self-dual or anti-self-dual,
so for general connections these are independent. For example, in
Eq.(\ref{Weyl integrability}) the three terms lie in different subspaces.
Clearly we may vary the Weyl vector independently of $\hat{\boldsymbol{\omega}}^{ac}$,
so $\gamma\sigma=0$. For the connection and its dual we imagine a
case with only $\hat{\boldsymbol{\omega}}^{01}$ nonzero. Then the
$01$ component $e_{\;\quad bd}^{01}\hat{\boldsymbol{\omega}}^{bd}$
vanishes and we require $\gamma\rho$ and $\gamma\lambda$ to vanish
independently. 

Next, look at integrability of the modified $\mathbf{f}_{a}$, which
changes through its dependence on the other fields.
\[
0=\mathbf{d}^{2}\mathbf{f}_{a}=\mathbf{d}\boldsymbol{\omega}_{\;\;\;a}^{b}\wedge\mathbf{f}_{b}-\boldsymbol{\omega}_{\;\;\;a}^{b}\wedge\mathbf{d}\mathbf{f}_{b}+\mathbf{d}\mathbf{f}_{a}\wedge\boldsymbol{\omega}-\mathbf{f}_{a}\wedge\mathbf{d}\boldsymbol{\omega}
\]
Substituting Eqs.(\ref{Extended connection},\ref{Extended f}, and
\ref{Extended Weyl}) and collecting terms we find
\begin{eqnarray*}
0 & = & \left(\beta+\gamma\right)\mathbf{e}^{b}\land\mathbf{f}_{a}\wedge\mathbf{f}_{b}
\end{eqnarray*}
so we must set $\beta=-\gamma$.

Insuring integrability of the solder form, $\mathbf{d}^{2}\mathbf{e}^{a}=0$,
is longer. The same procedure eventually leads to
\begin{eqnarray*}
0 & = & -\rho\boldsymbol{\omega}^{ac}\land\boldsymbol{\omega}_{\;\;\;c}^{e}\wedge\mathbf{f}_{e}\\
 &  & +\sigma\left(\mathbf{d}g^{ac}+\sigma\boldsymbol{\omega}^{ac}+\sigma\boldsymbol{\omega}^{ca}\right)\wedge\boldsymbol{\omega}\land\mathbf{f}_{c}-2\rho\boldsymbol{\omega}^{ac}\land\mathbf{f}_{c}\wedge\boldsymbol{\omega}\\
 &  & +\left(\gamma\rho-2\mu-\gamma\sigma-2\alpha\lambda\right)g^{ae}\mathbf{e}^{c}\land\mathbf{f}_{e}\land\mathbf{f}_{c}\\
 &  & -\lambda\left(e_{\;\quad de}^{bc}\boldsymbol{\omega}^{de}\wedge\boldsymbol{\omega}_{\;\;\;b}^{a}\land\mathbf{f}_{c}-e_{\;\quad bd}^{ac}\boldsymbol{\omega}^{de}\wedge\boldsymbol{\omega}_{\;\;\;e}^{b}\land\mathbf{f}_{c}-e_{\;\quad bd}^{ac}\boldsymbol{\omega}^{bd}\land\boldsymbol{\omega}_{\;\;\;c}^{e}\wedge\mathbf{f}_{e}\right)\\
 &  & -2\lambda e_{\;\quad bd}^{ac}\boldsymbol{\omega}\wedge\boldsymbol{\omega}^{bd}\land\mathbf{f}_{c}\\
 &  & -\alpha e_{\;\;\;b\;\;\quad d}^{a\;\;\quad c}\mathbf{e}^{b}\wedge\mathbf{e}^{d}\land\mathbf{f}_{c}\\
 &  & +\left(2\beta\lambda-\alpha\rho\right)e_{\;\;\;\quad d}^{ace}\mathbf{f}_{c}\land\mathbf{f}_{e}\land\mathbf{e}^{d}\\
 &  & -2\mu\lambda e^{acfe}\mathbf{f}_{e}\land\mathbf{f}_{f}\land\mathbf{f}_{c}
\end{eqnarray*}
where each row is independent. In particular, the singleton terms
show that $\rho=\lambda=\alpha=0$. Dropping these coefficients reduces
the integrability condition to
\begin{eqnarray*}
0 & = & \sigma\left(\mathbf{d}g^{ac}+\sigma\boldsymbol{\omega}^{ac}+\sigma\boldsymbol{\omega}^{ca}\right)\wedge\boldsymbol{\omega}\land\mathbf{f}_{c}\\
 &  & -2\left(\mu+\frac{1}{2}\gamma\sigma\right)g^{ae}\mathbf{e}^{c}\land\mathbf{f}_{e}\land\mathbf{f}_{c}
\end{eqnarray*}
The first term is proportional to $\boldsymbol{\omega}\wedge\boldsymbol{\omega}\land\mathbf{f}_{c}=0$
and we are left with
\[
\mu=-\frac{1}{2}\gamma\sigma
\]
Integrability of the remaining structure equation is also long, but
with the conditions already identified all terms cancel identically,
$\mathbf{d}^{2}\boldsymbol{\omega}_{\;\;\;b}^{a}\equiv0$. We conclude
that the maximally extended Lie algebra is given by
\begin{eqnarray*}
\mathbf{d}\boldsymbol{\omega}_{\;\;\;b}^{a} & = & \boldsymbol{\omega}_{\;\;\;b}^{c}\wedge\boldsymbol{\omega}_{\;\;\;c}^{a}-\gamma\left(\delta_{b}^{c}\delta_{d}^{a}-\eta^{ac}\eta_{bd}\right)\mathbf{e}^{d}\land\mathbf{f}_{c}-\frac{1}{2}\gamma\sigma\left(\eta^{ad}\delta_{b}^{c}-\eta^{ac}\delta_{b}^{d}\right)\mathbf{f}_{d}\land\mathbf{f}_{c}\\
\mathbf{d}\mathbf{e}^{a} & = & \mathbf{e}^{b}\wedge\boldsymbol{\omega}_{\;\;\;b}^{a}+\boldsymbol{\omega}\wedge\mathbf{e}^{a}+\sigma\eta^{ac}\boldsymbol{\omega}\land\mathbf{f}_{c}\\
\mathbf{d}\mathbf{f}_{a} & = & \boldsymbol{\omega}_{\;\;\;a}^{b}\wedge\mathbf{f}_{b}+\mathbf{f}_{a}\wedge\boldsymbol{\omega}\\
\mathbf{d}\boldsymbol{\omega} & = & \gamma\mathbf{e}^{c}\land\mathbf{f}_{c}
\end{eqnarray*}
These take a more recognizable form with the substitutions
\begin{align*}
\mathbf{f}_{a} & =\frac{1}{\gamma}\tilde{\mathbf{f}}_{a}\\
\mathbf{e}^{a} & =\tilde{\mathbf{e}}^{a}-\frac{\sigma}{2\gamma}\eta^{ab}\tilde{\mathbf{f}}_{b}
\end{align*}
These describe the Lie algebra $\mathfrak{so}\left(p+1,q+1\right)$
of the conformal group,
\begin{eqnarray}
\mathbf{d}\boldsymbol{\omega}_{\;\;\;b}^{a} & = & \boldsymbol{\omega}_{\;\;\;b}^{c}\wedge\boldsymbol{\omega}_{\;\;\;c}^{a}-\left(\delta_{b}^{c}\delta_{d}^{a}-\eta^{ac}\eta_{bd}\right)\tilde{\mathbf{e}}^{d}\land\tilde{\mathbf{f}}_{c}\nonumber \\
\mathbf{d}\tilde{\mathbf{e}}^{a} & = & \tilde{\mathbf{e}}^{b}\wedge\boldsymbol{\omega}_{\;\;\;b}^{a}+\boldsymbol{\omega}\wedge\tilde{\mathbf{e}}^{a}\nonumber \\
\mathbf{d}\tilde{\mathbf{f}}_{a} & = & \boldsymbol{\omega}_{\;\;\;a}^{b}\wedge\tilde{\mathbf{f}}_{b}+\tilde{\mathbf{f}}_{a}\wedge\boldsymbol{\omega}\nonumber \\
\mathbf{d}\boldsymbol{\omega} & = & \tilde{\mathbf{e}}^{a}\land\tilde{\mathbf{f}}_{a}\label{Conformal Maurer Cartan}
\end{eqnarray}
Here we take the conformal weight of the solder form to be $+1$ so
that the metric $g_{\alpha\beta}=e_{\alpha}^{\;\;\;a}e_{\beta}^{\;\;\;b}\eta_{ab}$
has weight $+2$. It follows that $\eta_{ab}$ has conformal weight
zero and satisfies $\mathbf{d}\eta_{ab}=0$.

Equations (\ref{Conformal Maurer Cartan}) are the Maurer-Cartan structure
equations of the conformal group. This makes rigorous the conjecture
of \cite{Wheeler 2023b}, that the mixed symmetry nonmetricity is
related to conformal symmetry. This equivalence becomes more striking
when we modify to include curvatures.

\section{Developing the curvatures \label{sec:Developing-the-curvatures}}

To develop the curvatures, we return to the $\mathbf{u}^{a},\mathbf{v}_{a}$
basis. In this basis the torsion and nonmetricity separate.

We express the conformal equations (\ref{Conformal Maurer Cartan})
in terms of $\mathbf{u}^{a}$ and $\mathbf{v}_{a}$ by inverting Eqs.(\ref{(u,v) in terms of (e,f) basis}).
This results in
\begin{eqnarray*}
\mathbf{d}\boldsymbol{\omega}_{\;\;\;b}^{a} & = & \boldsymbol{\omega}_{\;\;\;b}^{c}\wedge\boldsymbol{\omega}_{\;\;\;c}^{a}+2\left(g^{ae}\mathbf{v}_{e}\land\mathbf{v}_{b}-\mathbf{u}^{a}\land g_{bc}\mathbf{u}^{c}\right)\\
\mathbf{d}\mathbf{u}^{a} & = & \mathbf{u}^{b}\wedge\boldsymbol{\omega}_{\;\;\;b}^{a}+\boldsymbol{\omega}\wedge\mathbf{u}^{a}\\
\mathbf{d}\mathbf{v}_{a} & = & \boldsymbol{\omega}_{\;\;\;a}^{b}\wedge\mathbf{v}_{b}+\mathbf{v}_{a}\wedge\boldsymbol{\omega}\\
\mathbf{d}\boldsymbol{\omega} & = & 2\mathbf{u}^{a}\land\mathbf{v}_{a}
\end{eqnarray*}
Here the structure equations for $\mathbf{u}^{a}$ and $\mathbf{v}_{a}$
are unchanged from the original separation of torsion and nonmetricity
given by Eqs.(\ref{Cartan Eqs so far}). Therefore, when we restore
curvatures we retain the previous relations for $\mathbf{d}\mathbf{u}^{a}$
and $\mathbf{d}\mathbf{v}_{a}$, giving
\begin{eqnarray*}
\mathbf{d}\boldsymbol{\omega}_{\;\;\;b}^{a} & = & \boldsymbol{\omega}_{\;\;\;b}^{c}\wedge\boldsymbol{\omega}_{\;\;\;c}^{a}+2\left(\mathbf{u}^{a}\land g_{bc}\mathbf{u}^{c}-g^{ac}\mathbf{v}_{c}\land\mathbf{v}_{b}\right)+\mathbf{R}_{\;\;\;b}^{a}\\
\mathbf{d}\mathbf{u}^{a} & = & \mathbf{u}^{b}\wedge\boldsymbol{\omega}_{\;\;\;b}^{a}+\boldsymbol{\omega}\wedge\mathbf{u}^{a}+\boldsymbol{\mathcal{T}}^{a}\\
\mathbf{d}\mathbf{v}_{a} & = & \boldsymbol{\omega}_{\;\;\;a}^{b}\wedge\mathbf{v}_{b}+\mathbf{v}_{a}\wedge\boldsymbol{\omega}+\tilde{\mathbf{Q}}_{a}\\
\mathbf{d}\boldsymbol{\omega} & = & 2\mathbf{u}^{a}\wedge\mathbf{v}_{a}+\boldsymbol{\Omega}
\end{eqnarray*}
Here $\mathbf{R}_{\;\;\;b}^{a}$ and $\boldsymbol{\Omega}$ are to
be determined, but $\boldsymbol{\mathcal{T}}^{a}$ and $\tilde{\mathbf{Q}}_{a}$
are the original fields.

Next, to relate the torsion and nonmetricity to conformal curvatures
we invert to transform back to the conformal frame $\left(\mathbf{e}^{a},\mathbf{f}_{a}\right)$,
with $\mathbf{v}_{a}=\frac{1}{2}\left(\mathbf{f}_{a}+g_{ab}\mathbf{e}^{b}\right)$
and $\mathbf{u}^{a}=\frac{1}{2}\left(\mathbf{e}^{a}-g^{ab}\mathbf{f}_{b}\right)$
to find
\begin{eqnarray}
\mathbf{d}\boldsymbol{\omega}_{\;\;\;b}^{a} & = & \boldsymbol{\omega}_{\;\;\;b}^{c}\wedge\boldsymbol{\omega}_{\;\;\;c}^{a}+\left(\delta_{c}^{a}\delta_{b}^{d}-g^{ad}g_{bc}\right)\mathbf{f}_{d}\land\mathbf{e}^{c}+\mathbf{R}_{\;\;\;b}^{a}\nonumber \\
\mathbf{d}\mathbf{e}^{a} & = & \mathbf{e}^{b}\wedge\boldsymbol{\omega}_{\;\;\;b}^{a}+\boldsymbol{\omega}\wedge\mathbf{e}^{a}+\mathbf{T}^{a}\nonumber \\
\mathbf{d}\mathbf{f}_{a} & = & \boldsymbol{\omega}_{\;\;\;a}^{b}\wedge\mathbf{f}_{b}+\mathbf{f}_{a}\wedge\boldsymbol{\omega}+\bar{\mathbf{T}}_{a}\nonumber \\
\mathbf{d}\boldsymbol{\omega} & = & \mathbf{e}^{a}\wedge\mathbf{f}_{a}+\boldsymbol{\Omega}\label{Gauged conformal group}
\end{eqnarray}
with $\mathbf{T}^{a},\bar{\mathbf{T}}_{a}$ as defined in Section
(\ref{sec:Restoring-independence-of}). This is precisely the form
of the gauged conformal group, with torsion $\mathbf{T}^{a}$ and
special conformal field strength $\bar{\mathbf{T}}_{a}$. The evident
parallel structure of the solder form and special conformal transformations,
$\boldsymbol{\mathcal{T}}^{a}\pm\tilde{\mathbf{Q}}_{a}$, emphasizes
the understanding of mixed symmetry nonmetricity $\tilde{\mathbf{Q}}_{a}$
as torsion-like.

Bringing the calculation full circle, we reconstruct the Cartan equations
with the asymmetric connection. Choose the orthonormal metric so that
$\mathbf{d}\eta_{ab}=0$. Replacing $\boldsymbol{\omega}_{\;\;\;b}^{a}=\hat{\boldsymbol{\omega}}_{\;\;\;b}^{a}+\frac{1}{2}\left(\tilde{\mathbf{Q}}_{\;\;\;b}^{a}+2\delta_{b}^{a}\boldsymbol{\omega}\right)$
we find the maximal nonmetric system
\begin{eqnarray}
\mathbf{d}\hat{\boldsymbol{\omega}}_{\;\;\;b}^{a} & = & \hat{\boldsymbol{\omega}}_{\;\;\;b}^{c}\wedge\hat{\boldsymbol{\omega}}_{\;\;\;c}^{a}+\left(\delta_{c}^{a}\delta_{b}^{d}-\eta^{ad}\eta_{bc}\right)\mathbf{f}_{d}\land\mathbf{e}^{c}+\boldsymbol{\mathcal{R}}_{\;\;\;b}^{a}\nonumber \\
\mathbf{d}\mathbf{e}^{a} & = & \mathbf{e}^{b}\wedge\hat{\boldsymbol{\omega}}_{\;\;\;b}^{a}+\boldsymbol{\mathcal{T}}^{a}\nonumber \\
\mathbf{d}\mathbf{f}_{a} & = & \hat{\boldsymbol{\omega}}_{\;\;\;a}^{b}\wedge\mathbf{f}_{b}+\boldsymbol{\mathcal{K}}_{a}\nonumber \\
\mathbf{d}\boldsymbol{\omega} & = & \mathbf{e}^{a}\wedge\mathbf{f}_{a}+\boldsymbol{\Omega}\label{Maximal nonmetric system}
\end{eqnarray}
where
\begin{eqnarray*}
\boldsymbol{\mathcal{R}}_{\;\;\;b}^{a} & = & \mathbf{R}_{\;\;\;b}^{a}+\frac{1}{2}\mathbf{D}\hat{\mathbf{Q}}_{\;\;\;b}^{a}-\frac{1}{4}\hat{\mathbf{Q}}_{\;\;\;b}^{c}\wedge\hat{\mathbf{Q}}_{\;\;\;c}^{a}\\
\boldsymbol{\mathcal{K}}_{a} & = & \boldsymbol{\mathcal{T}}_{a}+\frac{1}{2}\left(\mathbf{e}^{b}-g^{bc}\mathbf{f}_{c}\right)\wedge\tilde{\mathbf{Q}}_{ab}
\end{eqnarray*}
Eqs.(\ref{Maximal nonmetric system}) reduce to the original system
when $\mathbf{f}_{a}=\mathbf{h}_{a}=\eta_{ab}\mathbf{e}^{b}$.

Therefore, \emph{the Maurer-Cartan equations of the maximally extended
general linear system may be recast as the Maurer-Cartan equations
of conformal symmetry, with the nonmetricity of the asymmetric system
equaling the field strength of special conformal transformations in
the conformal system.}

In the general linear frame, the curvature has a symmetric part. Writing
$\mathbf{d}\hat{\boldsymbol{\omega}}^{ab}+\mathbf{d}\hat{\boldsymbol{\omega}}^{ba}$
to symmetrize the equation, then expanding the connection in terms
of antisymmetric and symmetric parts, $\boldsymbol{\omega}^{ca}+\frac{1}{2}\left(\hat{\boldsymbol{\omega}}^{ac}+\hat{\boldsymbol{\omega}}^{ca}\right)$
we find that $-\frac{1}{2}\mathbf{D}\hat{\mathbf{Q}}_{\;\;\;b}^{a}+\frac{1}{4}\hat{\mathbf{Q}}_{\;\;\;b}^{c}\wedge\hat{\mathbf{Q}}_{\;\;\;c}^{a}$
cancels identically leaving 
\begin{eqnarray*}
\mathbf{R}^{ab}+\mathbf{R}^{ba} & = & 0
\end{eqnarray*}
This shows that the Lorentzian curvature is properly antisymmetric
in Eqs.(\ref{Gauged conformal group}).

Notice that if the symmetric part of the curvature $\boldsymbol{\Omega}^{ab}+\boldsymbol{\Omega}^{ba}=\mathbf{D}\mathbf{Q}^{ab}$
vanishes then it follows that
\begin{eqnarray*}
0 & = & \eta_{ab}\mathbf{D}\mathbf{Q}^{ab}\\
 & = & \mathbf{D}\left(\eta_{ab}\mathbf{Q}^{ab}\right)-\mathbf{D}\eta_{ab}\wedge\mathbf{Q}^{ab}\\
 & = & 2n\,\mathbf{d}\boldsymbol{\omega}-\mathbf{Q}_{ab}\wedge\mathbf{Q}^{ab}
\end{eqnarray*}
and since $\mathbf{Q}_{ab}\wedge\mathbf{Q}^{ab}=0$ the Weyl vector
is integrable, $\mathbf{d}\boldsymbol{\omega}=0$.

\section{Discussion}

Curvature, torsion, and nonmetricity have distinct physical effects.
The geodesic deviation described by curvature is familiar, and it
is not hard to see that torsion can produce anomalous precession of
angular momentum beyond the Lense-Thirring effect. By contrast to
these, nonmetricity affects the parallel transport of angles, so that
two identical cubes parallel transported along different paths become
non-identical parallelepipeds when later compared. It is therefore
surprising that there is a direct equivalence between a substantial
part of the nonmetricity and the torsion.

In demonstrating these relationships we considered three systems,
equivalent via field redefinitions. Summarizing each by its independent
variables and corresponding curvatures or field strengths, we have
the nonmetric $\left(NM\right)$, Lorentzian with separated sources
$\left(L,sep\right)$ and the conformal $\left(C\right)$ :
\begin{equation}
\left[\begin{array}{cc}
\hat{\boldsymbol{\omega}}_{\;\;\;b}^{a} & \boldsymbol{\mathcal{R}}_{\;\;\;b}^{a}\\
\mathbf{e}^{a} & \boldsymbol{\mathcal{T}}^{a}\\
\mathbf{f}_{a} & \tilde{\mathbf{Q}}_{\;\;\;b}^{a}\\
\boldsymbol{\omega} & \boldsymbol{\Omega}
\end{array}\right]_{NM}\cong\left[\begin{array}{cc}
\boldsymbol{\omega}_{\;\;\;b}^{a} & \mathbf{R}_{\;\;\;b}^{a}\\
\mathbf{u}^{a} & \boldsymbol{\mathcal{T}}^{a}\\
\mathbf{v}_{a} & \tilde{\mathbf{Q}}_{a}\\
\boldsymbol{\omega} & \boldsymbol{\Omega}
\end{array}\right]_{L,sep}\cong\left[\begin{array}{cc}
\boldsymbol{\omega}_{\;\;\;b}^{a} & \mathbf{R}_{\;\;\;b}^{a}\\
\mathbf{e}^{a} & \boldsymbol{\mathcal{T}}^{a}-\tilde{\mathbf{Q}}^{a}\\
\mathbf{f}_{a} & \boldsymbol{\mathcal{T}}_{a}+\tilde{\mathbf{Q}}_{a}\\
\boldsymbol{\omega} & \boldsymbol{\Omega}
\end{array}\right]_{C}\label{3 systems}
\end{equation}

There are two striking results here. Firstly, symmetric contributions
to the connection due to nonmetricity may be completely absorbed by
field redefinitions to restore the antisymmetric spin connection of
a Lorentzian system\footnote{This is one of those results that, if you stare at it long enough,
seems obvious--naturally we can always choose an orthonormal basis
to reduce a general linear system to an orthonormal (Lorentzian) one
(see Theorem 5.8 in \cite{Isham}). But here this correspondence drops
effortlessly out of the structure equations, and applies with Lorentz
symmetry.}. Secondly, the field strength $\mathbf{D}\mathbf{f}_{a}$ of the
special conformal gauge field $\mathbf{f}_{a}$ equals the sum of
the original torsion and nonmetricity.
\[
\left.\mathbf{D}\mathbf{f}_{a}\right|_{Conformal\,basis}=\left.\boldsymbol{\mathcal{T}}_{a}+\tilde{\mathbf{Q}}_{a}\right|_{Asymmetric\,basis}
\]
This is an unexpected property of the special conformal transformations\footnote{It does not matter how long we stare.}.

The set of Cartan equations (\ref{Gauged conformal group}) describe
the usual conformal gauge theory. This form of $SO\left(p+1,q+1\right)$
gauge theory has a long history (\cite{Weyl 1918a}-\cite{HobsonLazenby})
and multiple interpretations, depending on our choice of fiber bundle.
We remark on the two interpretations of $\mathbf{f}_{a}=\mathbf{d}y_{\alpha}$
identified in Subsection (\ref{subsec:Integrability-of-nonmetric}):
\begin{itemize}
\item Auxiliary conformal gravity is based locally on the quotient of the
conformal group by its inhomogeneous Weyl subgroup, $\mathcal{M}^{d}=\mathcal{C}/\mathcal{IW}$,
on a $d$-dimensional manifold. This corresponds to the first case,
$\mathbf{f}_{a}=b_{ab}\left(x^{c}\right)\mathbf{e}^{b}$. When the
torsion $\mathbf{S}^{a}$ vanishes the special conformal gauge field
equals the Schouten tensor, 
\[
\mathbf{f}_{a}=\mathscr{R}_{ab}\mathbf{e}^{b}=\frac{1}{n-2}\left(R_{ab}-\frac{1}{2\left(n-1\right)}\eta_{ab}R\right)\mathbf{e}^{b}
\]
This solution enforces conformal structure by making $\boldsymbol{\Omega}_{\;\;\;b}^{a}$
equal to the Weyl curvature $\mathbf{C}_{\;\;\;b}^{a}$, regardless
of the particular action. In an Einstein space, $R_{ab}-\frac{1}{2}g_{ab}R=\Lambda g_{ab}$
we have 
\begin{eqnarray*}
\mathscr{R}_{ab} & = & -\frac{1}{6}\Lambda g_{ab}
\end{eqnarray*}
and the field strength $\mathbf{D}\boldsymbol{\mathscr{R}}_{ab}$
of the special conformal gauge field $\mathbf{f}_{a}$ is exactly
proportional to the nonmetricity $\mathbf{D}\boldsymbol{\mathscr{R}}_{ab}=-\frac{1}{6}\Lambda\mathbf{Q}_{ab}$.
\item Biconformal gravity \cite{Wheeler1998,Wheeler1999} is based on the
quotient of the conformal group by the homogeneous Weyl group, $\mathcal{M}^{2d}=\mathcal{C}/\mathcal{W}$.
The quotient is a Kähler manifold, with $y_{a}$ independent of $x^{a}$.
The volume element is dimensionless and it becomes possible to write
a scale invariant curvature-linear action in any dimension. With vanishing
torsion $\mathbf{S}_{a}$, the resulting gravity theory reduces to
scale covariant (i.e., integrable Weyl) general relativity on the
co-tangent bundle \cite{Wheeler2019b}. Here the torsion $\mathbf{T}_{a}$
and the co-torsion $\bar{\mathbf{T}}_{a}$ play equivalent roles as
torsions associated with translations at the antipodes of compactified
Minkowski space.
\end{itemize}
In conclusion, independent variation of the metric and connection
of general linear gravity theory leads to torsion and nonmetricity.
From this starting point we showed how field redefinitions reduce
the system to Poincaré gauge gravity. We showed that the maximal extension
of the Poincaré system to allow determination of both the original
torsion and the nonmetricity leads to a conformal geometry. The original
torsion $\boldsymbol{\mathcal{T}}_{a}$ and traceless, mixed-symmetry
$\tilde{\mathbf{Q}}^{a}$ of the general linear theory mix as the
torsion $\mathbf{T}^{a}=\boldsymbol{\mathcal{T}}_{a}-\tilde{\mathbf{Q}}^{a}$
and special conformal field strength $\bar{\mathbf{T}}^{a}=\boldsymbol{\mathcal{T}}_{a}+\tilde{\mathbf{Q}}^{a}$
of the conformal system. In the course of our proof, we found the
three equivalent systems (\ref{3 systems}) of Cartan equations.

The nonmetricity is comprised of irreducible mixed-symmetry $\hat{\mathbf{Q}}^{a}$
and totally symmetric $Q_{\left(abc\right)}$ parts, with the latter
driven only by Spin-3 sources or terms higher than quadratic in $Q_{\left(abc\right)}$
in the action. In the absence of consistent Spin-3 fields and the
equivalence of $\hat{\mathbf{Q}}^{a}$ to a combination of conformal
fields, we conclude that the study of gravity theories with general
connections may be recast as a study of conformal gauge theories of
gravity.

\end{document}